# MICROMAGNETIC MODELING OF MAGNETIZATION DISTRIBUTION CAUSED BY INTERNAL TWIN MICROSTRUCTURE OF NiMnGa FERROMAGNETIC SHAPE MEMORY ALLOYS

A.A. Likhachev and Yu.N. Koval

*Institute for Metal Physics, National Academy of Sciences, 36 Vernadsky Str., 03680,Kiev, Ukraine,* e-mail: alexl@imp.kiev.ua

**PACS:** 81.30.Kf; 62.20.fg; 61.72.Mm; 75.30.Gw; 75.60.Ch; 75.60.Ej

**Keywords:** martensitic transformations; shape-memory effect; twinning; magnetic anisotropy; domain walls and domain structure; magnetization curves.

## Abstract

In our present publication we will continue studying the effects of the magnetostatic energy on the magnetization behavior of FMSMA's recently started in some our publications. Our method is based on the direct minimization of our new micromagnetic free energy model of FMSMA's taking into account both the magnetic anisotropy energy and the magnetostatic energy contributions associated with the laminated two-variant twin microstructure. Here, our special interest is to investigate a distribution of local magnetizations within both twin variants, their orientation and their dependence on the external magnetic field and their volume fractions, as well. We will use a direct optimization method based on the the Nelder–Mead algorithm which is applied to find the minimum of our model magnetic free energy function. We will compare and discuss the obtained results with some earlier models.

## Introduction

Since 1996 Ni-Mn-Ga based Ferromagnetic Shape Memory Alloys (FMSMA's) were considered as best candidates having a unical ability to show an extremely large magnetic field induced deformation effects which is about of 30-50 times larger compared to the best known ordinary magnetostrictive materials [1]. First, these effects were discovered experimentally in two different nonstochiometric ferromagnetic martensitic phases of NiMnGa alloy ( 6% in 5M [2] and then 10% in 7M) [3]. Sometime ago, even larger 12% magnetic field induced strain (MFIS) effect was obtained in Ni46Mn24Ga22Co4Cu4 non-modulated (NM) martensite [22]. It has been found that the strain mechanism in FMSMA's is based on the twin boundary motion and the resulting redistribution between two twin related variants A and B of the martensitic phase, which easy magnetization axes are perpendicular each to other [8-12, 17-21]. Both the multiple twin microstructure and the magneto-optical image of $180^0$ magnetic domains inside of the twin bands of the 5M martensite of NiMnGa alloy are shown in Fig.1 below.

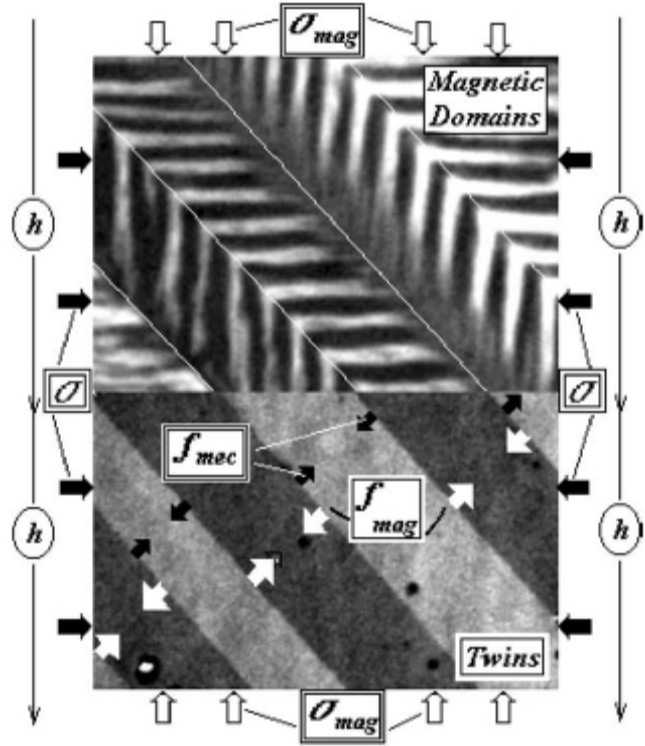

**Fig.1**. The twin microstructure consisting of two martensite variants A and B and the magneto-optical image of $180^0$ magnetic domains within the internally twinned 5M martensite of NiMnGa alloy.

It has been proved that the twinning process in NiMnGa is driven by a macroscopic magnetostrictive force developed during the magnetization of this material.

Generally, the magnetic forces can be produced in any ferromagnetic material when it magnetizes. Everything depends on the fact: if such material is deformable and if its magnetic free energy per unit volume $F^{mag}(\mathbf{h},\boldsymbol{\varepsilon})$ is dependent not only on the external magnetic field $\mathbf{h}$, but also on the strain of the material $\boldsymbol{\varepsilon}$. Here, we have defined the magnetic free energy to be zero at $\mathbf{h}=0$. So, it represents only a magnetic part of the total free energy of the material. In such a case both the macroscopic magnetization of the material $\mathbf{M}(\mathbf{h},\boldsymbol{\varepsilon})$ and the field-induced magnetic forces $\boldsymbol{\sigma}^{mag}(\mathbf{h},\boldsymbol{\varepsilon})$ can be represented on the basis of the general thermodynamic relationships as follows:

$$\mu_0 \mathbf{M}(\mathbf{h},\boldsymbol{\varepsilon}) = -\left(\frac{\partial}{\partial \mathbf{h}} F^{mag}(\mathbf{h},\boldsymbol{\varepsilon})\right)_{\boldsymbol{\varepsilon}} \qquad \boldsymbol{\sigma}^{mag}(\mathbf{h},\boldsymbol{\varepsilon}) = -\left(\frac{\partial}{\partial \boldsymbol{\varepsilon}} F^{mag}(\mathbf{h},\boldsymbol{\varepsilon})\right)_{\mathbf{h}} \qquad (1)$$

Here, $\mu_0 = 4\pi/10^7$ is the vacuum magnetic permeability. In case of FMSMA's the strain occurs due to the twin boundary motion shown in Fig.1 by white arrows. So, the (x, y) strain components will change proportionally to the twin variant fraction change:

$$\varepsilon_{yy} = -\varepsilon_{xx} = \varepsilon_0 x_A; \qquad \varepsilon_{zz} = 0; \qquad (2)$$

Here and after $x_A$ and $x_B$ will denote the volume fractions of both twin variants A and B, correspondingly.

According to [8], this force can be generally defined as follows:

$$\sigma_{mag}(h, x_A) = -\frac{\partial}{\partial \varepsilon} F_{mag}(h, x_A); \quad \text{where}, \quad \varepsilon = \varepsilon_0 x_A \tag{3}$$

Here, $\sigma_{mag}(h, x_A)$ is the magnetostrictive force component acting along the external magnetic field per unit cross-section ( as shown in **Fig.1** ). Here, $F_{mag}(h, x_A)$ is the magnetic free energy per unit volume dependent on the magnetic field $h$ and the volume fraction $x_A$ occupied by the martensite variant A. The large strain $\varepsilon = \varepsilon_0 x_A$ which is developed in FMSMA's increases proportionally to the volume fraction $x_A$ and achieves its maximal value $\varepsilon_0$, which is a crystallographic constant dependent on the martensitic crystal lattice parameters. In NiMnGa alloys it can take 6% in 5M and 10% in 7M martensitic phases [8-12] or 12% in NM phase [22], correspondingly. In particular, both the magnetic field induced strain and the volume fractions can be found as a solution of the general force balance equation [10, 23]:

$$\sigma_{mag}(h, x_A) = \sigma + \sigma_0(\varepsilon); \quad \text{where}, \quad \varepsilon = \varepsilon_0 x_A \tag{4}$$

when both the magnetic field $h$ and the external bias mechanical stress $\sigma$ (shown in Fig.1 with black arrows) are applied simultaniously. Here, $\sigma_0(\varepsilon)$ represents a zero field stress-strain loading- unloading relationships that can be obtained from the mechanical testing of FMSMA's.

As follows from the previous discussion, finding the magnetic free energy of FMSMA's and its dependence on the magnetic field and the volume fractions of twin variants is an important task for understanding of these materials. An overview about the different theoretical approaches and their analysis one can find in [15]. To overcome some disadvantages these early models we recently proposed a new micromagnetic theory where along with the magnetic anisotropy and Zeeman's energies, the magnetostatic energy also plays an important role and produces a special coupling effect between the twin related martensite variants [13, 14]. It is caused by the demagnetizing effects both on the surface of the MSM material and also on the internal twin boundaries. In our present publication we will continue studying the effects of the magnetostatic energy on the magnetization behavior of FMSMA's recently started in [16]. Here, our special interest is to investigate a distribution of local magnetizations within both twin variants A and B, their orientation and their dependence on the external magnetic field and volume fractions $x_A$ and $x_B$, as well. We will use a direct optimization method based on the the Nelder–Mead algorithm (Nelder, J.A. and Mead, R. (1965)) which is a commonly applied numerical method used to find the minimum or maximum of a function in a multidimensional space. We will compare and discuss the obtained results with some earlier models.

**Magnetic free energy model**

As follows from our micromagnetic free energy model of FMSMA's like NiMnGa, proposed in [13, 14] and in a little bit different but equivalent form in [16], it consists of tree terms. The magnetic anisotropy energy is the first one:

$$F_{Ani+Zee} = K_U \left( x_A (\mathbf{m}_{A\perp}/M_s)^2 + x_B (\mathbf{m}_{B\perp}/M_s)^2 \right) - \mu_0 \mathbf{M}\mathbf{h} \tag{5}$$

Here, $\mu_0 = 4\pi/10^7$ is the vacuum magnetic permeability, what means that all the magnetic quantities will be represented in SI units instead of Gaussian system of units we used in [13, 14, 16] before. Here and later, $x_A$ and $x_B$ are the volume fractions of the twin related martensite variants A and B, respectively; $K_U$ and $M_s$ are the uniaxial magnetic anisotropy constant and saturation magnetization of the MSM material, correspondingly. Typically, $K_U \approx 1.7 \bullet 10^5 \, J/m^3$ and $M_s = 0.65T$ in NiMnGa-based FMSMA's. Everywhere below, $\mathbf{m}_A$ and $\mathbf{m}_B$ denote the local magnetizations averaged over the fine magnetic domain microstructure within both twin variants A and B. The subscript sign "$\perp$" means a magnetization component perpendicular to the local easy magnetization direction of the corresponding twin variant.

The second term in Eq.5 is a so-called Zeeman's energy describing the effect of the external magnetic field. It is dependent on the total magnetization $\mathbf{M} = x_A \mathbf{m}_A + x_B \mathbf{m}_B$ averaged over the twin microstructure and on the external magnetic field $\mathbf{h}$.

The third term represents the magnetostatic energy per unit volume:

$$U_{mag} = -\frac{1}{2}\mu_0 (x_A \mathbf{m}_A \mathbf{h}_A + x_B \mathbf{m}_B \mathbf{h}_B) \tag{6}$$

It also depends on the local demagnetizing field values $(\mathbf{h}_A)$ and $(\mathbf{h}_B)$ within the twin bands A and B, respectively. Similar to the macroscopic magnetization value $\mathbf{M} = x_A \mathbf{m}_A + x_B \mathbf{m}_B$ we can introduce the macroscopic demagnetizing field $\mathbf{H}^D = x_A \mathbf{h}_A + x_B \mathbf{h}_B$. Then, using identities $x_A = 1 - x_B$ and $x_B = 1 - x_A$ we can obtain:

$$\mathbf{h}_A = \mathbf{H}^D + x_B (\mathbf{h}_A - \mathbf{h}_B) \quad \text{and} \quad \mathbf{h}_B = \mathbf{H}^D - x_A (\mathbf{h}_A - \mathbf{h}_B) \tag{7}$$

According to a well known from the magnetism theory boundary conditions, both the normal component of the magnetic induction and the tangential components of the magnetic field must be continuous at the twin boundary interfaces. So, $(\mathbf{h}_A + \mathbf{m}_A)_\mathbf{n} = (\mathbf{h}_B + \mathbf{m}_B)_\mathbf{n}$ and $(\mathbf{h}_A)_\mathbf{t} = (\mathbf{h}_B)_\mathbf{t}$, respectively. Finally, these boundary conditions give us an important linear relationship between the local demagnetizing field and magnetization jumps at the twin boundaries:

$$\mathbf{h}_A - \mathbf{h}_B = -\mathbf{n}(\mathbf{m}_A - \mathbf{m}_B)_\mathbf{n} \tag{8}$$

Here $\mathbf{n}$ is a unit normal vector at the twin boundaries oriented at $45^0$ to the field direction: $n_x = n_y = 1/\sqrt{2}$; $n_z = 0$. There is also a well known linear relationship between the macroscopic demagnetizing field and the average magnetization of any ferromagnetic material:

$$\mathbf{H}^D = -\hat{\mathbf{N}}\mathbf{M} \tag{9}$$

Here, $\hat{\mathbf{N}}$ is a so-called demagnetizing matrix dependent only on the shape of a particular ferromagnetic sample. This matrix is always positively defined and has the unit its spur value. In a particular case, if the sample's shape is symmetric with respect to all reflections and inversion of **x**, **y**, and **z** coordinates in **Fig.1**, then a corresponding demagnetizing matrix will be a diagonal one with all positive matrix elements satisfying the unit spur value: $Sp\,\hat{\mathbf{N}} = D_{xx} + D_{yy} + D_{zz} = 1$.

Finally, the magnetostatic energy can be represented as follows:

$$U_{Mag} = \frac{1}{2}\mu_0 \left( \mathbf{M}\hat{\mathbf{N}}\mathbf{M} + x_A x_B (\mathbf{m}_A - \mathbf{m}_B)_\mathbf{n}^2 \right) \tag{7}$$

$$\mathbf{M} = x_A \mathbf{m}_A + x_B \mathbf{m}_B \tag{8}$$

Therefore, the magnetostatic energy consists of two terms. The first one is caused by an interaction between the magnetic charges induced at the external sample interface. In similar way, the second one represents an interaction between the magnetic charges induced at the twin boundaries.

Finally the total magnetic free energy is:

$$F_{Mag}(\mathbf{m}_A, \mathbf{m}_B) = F_{Ani+Zee}(\mathbf{m}_A, \mathbf{m}_B) + U_{Mag}(\mathbf{m}_A, \mathbf{m}_B) \tag{9}$$

It should be minimized with respect to the local magnetization variables $\mathbf{m}_A$ and $\mathbf{m}_B$, satisfying some additional restrictions:

$$(\mathbf{m}_A)^2 \leq M_s^2; \quad and \quad (\mathbf{m}_B)^2 \leq M_s^2 \tag{10}$$

It is convenient to introduce four dimensionless magnetization vector components: $\mathbf{v} = (v_{Ax}, v_{Ay}, v_{Bx}, v_{By})$ instead of $\mathbf{m}_A$ and $\mathbf{m}_B$ as follows:

$$\mathbf{m}_A = (m_{Ax}, m_{Ay}) = M_s(v_{Ax}, v_{Ay}); \quad \mathbf{m}_B = (m_{Bx}, m_{By}) = M_s(v_{Bx}, v_{By}) \tag{11}$$

After that the magnetic free energy can also be represented in a dimensionless form:

$$F_{Mag}(\mathbf{m}_A, \mathbf{m}_B) = \mu_0 M_s^2 f_{Mag}(\mathbf{v}) \tag{12}$$

Here,

$$f_{Mag}(\mathbf{v}) = k_u \left( x_A v_{Ax}^2 + x_B v_{By}^2 \right) + \frac{1}{2} \left[ D_{xx} \left( x_A v_{Ax} + x_B v_{Bx} \right)^2 + D_{yy} \left( x_A v_{Ay} + x_B v_{By} \right)^2 \right] +$$
$$+ \frac{1}{4} \left[ x_A x_B \left( (v_{Ax} - v_{Bx}) + (v_{Ay} - v_{By}) \right)^2 \right] - h_0 \left( x_A v_{Ay} + x_B v_{By} \right) \quad , \quad (13)$$

Here, $h_0$ is a dimensionless parameter characterizing the external magnetic field: $h = M_s h_0$ with- $M_s = 0.65T$. One more material parameter- $k_u = K_U / \mu_0 M_s^2$ represents the dimensionless magnetic anisotropy energy constant. This dimensionless magnetization free energy must be minimized within the four-dimension region:

$$v_{Ax}^2 + v_{Ay}^2 \leq 1; \quad v_{Bx}^2 + v_{By}^2 \leq 1 \quad (14)$$

**Minimizing procedure and results**

In this section we will consider the minimization free energy problem in one particular case when the FMSMA sample has a thin cylindrical shape, which is magnetized parallel to its long axis. It's well known, that if the cylinder is much longer than its diameter then the demagnetizing factor along its y-axis becomes zero, so as two other components along x- and z-axes will be equal ½. We will also choose the material parameters for the saturation magnetization and the anisotropy constant typical for the 5M-martensitic phase of NiMnGa FMSMA samples. So, we should take $M_s = 0.65T$ and $2K_U / M_s = \mu_0 H_s = 0.66T$ and define a dimensionless anisotropy constant: $k_u = K_u / \mu_0 M_s^2 = 0.51$. Here, $\mu_0 H_s = 0.65T$ is a so called anisotropy field.

Generally, it is expected, that a magnetization process in FMSMA's consists from three stages. At the first one, the absolute values of both local magnetizations will increase, remaining less of their saturation magnetization values: $|\mathbf{m}_A| \leq M_s, |\mathbf{m}_B| \leq M_s$. The magnetic field $h$ will increase from its zero value until the variant A becomes first fully saturated $|\mathbf{m}_A| = M_s$ at some critical field value: $h = h_A^S$.

At the second stage the magnetization within the variant A will remain constant $|\mathbf{m}_A| = M_s$ and may change its value by rotation only. At the same time, the magnetization within the variant B will continue to increase remaining less of its saturation value: $|\mathbf{m}_B| \leq M_s$, as the magnetic field $h$ increases from $h = h_A^S$ until the variant B becomes also fully saturated $|\mathbf{m}_B| = M_s$ at some second critical field value: $h = h_B^S$.

At the final third stage, both absolute local magnetization values will be remaining constant: $|\mathbf{m}_A| = M_s, |\mathbf{m}_B| = M_s$, as the magnetic field $h$ increases from $h = h_B^S$. During this stage both the magnetizations will change their values by rotation until their directions will become completely parallel to the external magnetic field direction.

Unfortunately, in presence of the magnetostatic interaction between the twin variants a minimization procedure should be done in the complex four dimension space area, where $\mathbf{m}_A^2(h, x_A) \leq M_s^2$ and $\mathbf{m}_B^2(h, x_A) \leq M_s^2$. So, it can be done only by using some numerical methods.

Practically, one can use the Nelder–Mead method (Nelder, J.A. and Mead, R. (1965)) which is a commonly applied numerical method used to find the minimum or maximum of a function in a multidimensional space. It is a direct search method which can be applied to nonlinear optimization problems for which derivatives may not be known.

In our paper we have used the *fminsearch*() function from MATLAB 7.0.4 version and performed a minimization procedure for the dimensionless magnetic free energy function $f_{Mag}(\mathbf{v})$ given by Eq.13,14.

At the first stage: $|\mathbf{m}_A| \leq M_s, |\mathbf{m}_B| \leq M_s$, the minimization procedure is performed over all four dimensionless local magnetization variables: $\mathbf{v} = (v_{Ax}, v_{Ay}, v_{Bx}, v_{By})$.

At the second stage: $|\mathbf{m}_A| = M_s, |\mathbf{m}_B| \leq M_s$, we must represent $\mathbf{v} = (\sin\varphi_A, \cos\varphi_A, v_{Bx}, v_{By})$ where the angle $\varphi_A$ between the $\mathbf{v}_A$ and the external field direction is introduced. So, we have only three independent variables to minimize $f_{Mag}(\mathbf{v})$ in this case.

At the third stage $|\mathbf{m}_A| = M_s, |\mathbf{m}_B| = M_s$ we need to introduce one more angle $\varphi_B$ between the $\mathbf{v}_B$ and the external field direction and represent: $\mathbf{v} = (\sin\varphi_A, \cos\varphi_A, \sin\varphi_B, \cos\varphi_B)$. So, $f_{Mag}(\mathbf{v})$ will be dependent only of two variables in this case

Finally we obtained the local magnetizations: $\mathbf{m}_A(h, x_A)$, $\mathbf{m}_B(h, x_A)$, the magnetic free energy: $F_{Mag}(h, x_A)$, and also the macroscopic magnetization curves $\mathbf{M} = x_A \mathbf{m}_A + x_B \mathbf{m}_B$ represented in Fig.2.

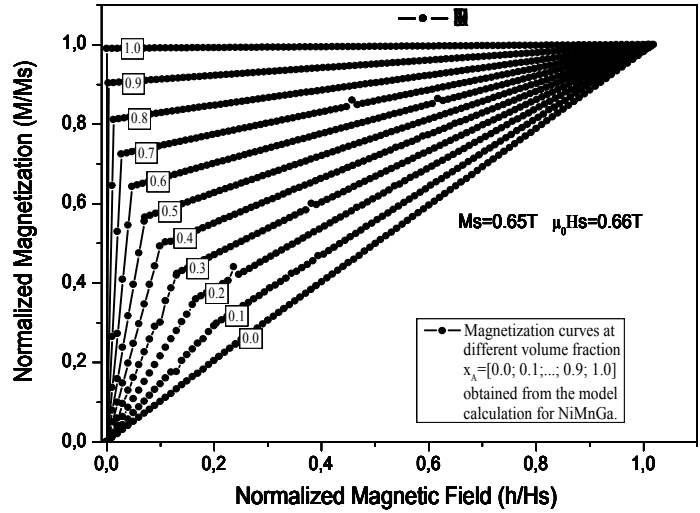

**Fig.2.** Magnetization curves obtained from the micromagnetic model calculations taking into account both the magnetic anisotropy and the magnetostatic energy contributions at zero demagnetizing factor along the magnetic field.

As follows from these results, at the first stage of the magnetization process both the local magnetizations of twin variants and the macroscopic magnetization must change linearly without change of their orientation until the full saturation in the variant A will happen at $h = h_A^S(x_A)$.

The saturation field $h = h_A^S(x_A)$ of the variant A is a singular point where, the magnetization curves have a jump-like change of their slope. Its value strongly depends on the volume fraction $x_A$. For the larger field values, at the second stage the magnetization curves show a nearly linear field dependence until a full saturation of the variant B will occur at $\mu_0 h = \mu_0 h_B^S = 2K_u / M_s$

For our calculations we have used the following material parameters typical for 5M martensitic phase of NiMnGa: the magnetic anisotropy constant $2K_u / M_s = 0.66T$, the saturation magnetization value $4\pi M_s = 0.65T$ and the demagnetizing factor along the field direction $D = 0$.

In our further publications these results will be used for the different theoretical calculations of the magnetic driving forces, magnetic field induced strain effects and some other interesting properties of FMSMA's.

**Acknowledgments** Present investigation has been supported from the National Budget Support Program No 6541230 "Support of the Favorite Scientific Direction Development" of Ukraine.

### References

[1] K. Ullakko, J. Mater. Eng. Perform. 5 (1996) 405-409.
[2] S.J. Murrey, M. Marioni, S.M. Allen, R.C. O'Handley, Appl. Phys. Lett. 77 (2000) 886-888.
[3] A. Sozinov, A.A. Likhachev, N. Lanska, K. Ullakko, Appl. Phys. Lett. 80 (2002) 1746-1748.

[4] K. Ullakko, J.K. Huang, C. Kantner, R.C. O'Handley, V.V. Kokorin, Appl. Phys. Lett. 69 (1996) 1966-1968.
[5] K. Ullakko, J.K. Huang, V.V. Kokorin, R.C. O'Handley, Scripta Mater. 36 (1997) 1133-1138.
[6] R.C. O'Handley, J. Appl. Phys. 83 (1998) 3263-3270.
[7] R.D. James, R.Tickle, M. Wuttig, Mater. Sci. Eng. A 273-275 (1999) 320-325.
[8] A.A. Likhachev, K. Ullakko, Eur. Phys. J. B 2 (1999) 1-9.
[9] A.A. Likhachev, K. Ullakko, Phys. Lett. A 275 (2000) 142-151.
[10] A.A. Likhachev, K. Ullakko, J. Phys. IV 11 (2001) Pr8-293-Pr8-298.
[11] A.A. Likhachev, A. Sozinov, K. Ullakko, Proc. SPIE 5387 (2004) 128-136.
[12] A.A. Likhachev, A. Sozinov, K. Ullakko, Mater. Sci. Eng. A 378(1-2) (2004) 513-518.
[13] A.A. Likhachev, Materials Science Forum Vols. 738-739 2013 pp 405-410.
[14] A.A. Likhachev, Chem. Met. Alloys, 2014, **6,** p.p.183-187.
[15] A.A. Likhachev, A. Sozinov, K. Ullakko, Mech. Mater. 38(5-6) (2006) 551-563.
[16] A.A. Likhachev and Yu.N. Koval, e-print, www. arXiv. org, 1912.02495 [cond-mat], 2019.
[17] O. Heczko, J. Magn. Magn. Mater. 290-291 (2005) 787-794.
[18] L. Straka, O. Heczko, J. Magn. Magn. Mater. 290-291 (2005) 829-831.
[19] U. Gaitzsch, H. Klauß, S. Roth, L. Schultz, J. Magn. Magn. Mater. 324 (2012) 430-433.
[20] Z. Li, Y. Zhang, C. Esling, X. Zhao, L. Zuo, Acta Mater. 59 (2011) 3390-3397.
[21] O. Heczko, J. Kopeček, L. Straka, H. Seiner, Mater. Res. Bull. 48 (2013) 5105-5109.
[22] A. Sozinov, N. Lanska, A. Soroka, W. Zou, Applied Physics Letters 102(2) ( 2013)/ DOI: 10.1063/1.4775677
[23] A.A. Likhachev, A. Sozinov, K. Ullakko, Proc. SPIE 4333 (2001) 197-206.